\documentstyle[amssymb,prl,aps,floats]{revtex}
\begin{document}
\twocolumn[
\hsize\textwidth\columnwidth\hsize\csname@twocolumnfalse\endcsname
\draft

\title{On the possibility of a double-well potential formation in diamond-like
amorphous carbon}
\author{A. Rakitin}
\address{Department of Electrical and Computer Engineering,
University of Toronto, 10 King's College Road, \\
Toronto, Ontario, Canada M5S 1A4}
\author{M. Ya. Valakh, N. I. Klyui, and V. G. Visotski}
\address{Institute of Semiconductor Physics,
National Academy of Sciences of Ukraine, 
Prospekt Nauki 45, 252650 Kiev-28, Ukraine }
\author {A. P. Litvinchuk}
\address{Texas Center for Superconductivity, University of Houston,
Houston, TX 77204-5932, USA}
\date{\today}
\maketitle

\begin{abstract}
A microscopic model, which describes specific features of the
electronic spectrum of various allotropic forms of amorphous carbon
as being responsible for  their structure peculiarities,  is presented.
It is shown that the formation of  a double-well potential is  a  
possible driving force for this behavior.
\end{abstract}

]

Amorphous carbon (a-C) and hydrogenated amorphous carbon (a-C:H) attract
attention not only due to their intriguing physical properties but also from
the point of view of their numerous application. Despite a large body of
theoretical and experimental research, the basic question of the
interrelationship of a-C atomic structure and its electronic properties
remains to be answered. The matter is that carbon has several allotropic
forms in contrast to other materials belonging to the fourth group such as, 
{\it e.g.}, Si and Ge characterized by tetrahedral fourfold sites ($sp^{3}$).
Carbon can adopt three different bonding configurations: $sp^{3}$, $sp^{2}$,
and $sp^{1}$. In the $sp^{3}$ configuration, each of the carbon's four
electrons is assigned to a tetrahedrally directed $sp^{3}$ hybrid orbital,
which forms a strong $\sigma $ bond with an adjacent atom. Such a
configuration is peculiar, {\it i.e.}, to diamond and corresponds to a
dielectric state with a wide energy gap of $5.5$ eV. At a carbon $sp^{2}$
site, three of the four electrons are assigned to the trigonally directed
$sp^{2}$ hybrids, which form $\sigma $ bonds, with the fourth electron
belonging to the $p_{z}(p\pi )$ orbital orthogonal to the $\sigma $ bonding
plane. This configuration is inherent to graphite, which is a good conductor
along its basal plane.

It is well known that Si and Ge atoms hold their tetrahedral coordination in
amorphous a-Si and a-Ge exhibiting a slight distortion in the bound angle
with respect to the value characteristic of a crystal. By contrast, there is
abundant evidence that amorphous a-C with the optical gap of $0.4-0.7$ eV is
predominantly $sp^{2}$ bonded [\ref{1} - \ref{5}]. This is not particularly
surprising, since graphite is precisely the allotrope of carbon which is
stable. This clearly demonstrates the importance of $\pi $-states in a-C as
they are responsible for the weakest inter-layer bonds closest to the Fermi
level on the energy scale. Thus they seem to affect the upper valence- and
conduction-band states to a great extent. There is a slight band overlap
($\simeq 0.04$ eV) in graphite with non-distorted $\pi $ states. As a
consequence, majority of theoretical models predict existence of only few
$sp^{3}$ bonded atoms in a-C. The necessity to consider the contribution both
from distorted $5$- and $7$-element rings [\ref{6},\ref{7}] and from twofold
carbon atoms [\ref{8},\ref{9}] was suggested in some theoretical models.

In the following we would like to discuss in more details a few models where
the electronic as well as structural properties of amorphous carbon received
primary emphasis. Beeman {\it et al}. [\ref{2}] considered three model
random networks of a-C with differing percentages of threefold- and
fourfold-coordinated atoms. The best agreement between theory and
experimental data on the radial distribution function was realized within
the model with $9\%$ fourfold coordinated atoms. Comparison of calculated
and experimental Raman scattering and vibrational density-of-states spectra
also suggests that the structure of amorphous carbon is composed of
three-coordinated planar regions and occasional four-coordinated atoms which
allow the plane orientation changes.

On the other hand, calculation of the electronic density-of-states function
by O'Reilly {\it et al}. [\ref{10}] carried out in the framework of Beeman's
model proved the absence of a gap at Fermi level. Based on this fact
Robertson [\ref{Robertson(Adv.Phys.1986)}] concluded that a-C possesses a
higher degree of $sp^{3}$ site ordering than it was assumed in
Ref. [\ref{2}]. The electronic structure of a-C was investigated by Robertson
and O'Reilly [\ref{4}]. They performed a simulation of a number of model
structures with different $sp^{2}-sp^{3}$ site ratio. This approach has led
to the cluster model that claimed $sp^{2}$ and $sp^{3}$ site segregation,
with $sp^{2}$ graphit clusters being embedded into $sp^{3}$-bounded matrix.
The gap $E_{g}$ of planar clusters varies with number of $6$-fold aromatic
rings $M$ as $E_{g}\propto 6/\sqrt{M}$. The typical $E_{g}\sim 0.5$ eV was
found to be consistent with clusters of about $15$ \AA\ in diameter. That
cluster approach was generally accepted for some years and was considered to
be a good starting point for the experimental data analysis. But recently
Robertson [\ref{7}] demonstrated that the cluster model [\ref{4}]
presupposes much more order and existence of large number of clasters than
can exist in a-C and a-C:H which are essentially disordered due to an ion
bombardment during deposition. In the context of cluster model it is further
difficult to explain the well-known experimental results on the optical gap
opening in completely $sp^{2}$-coordinated carbon (both in graphite and in
glassy-carbon) which is caused by an ion beam exposure [\ref{Compagnini}].
Compagnini {\it et al. [}\ref{Compagnini}{\it ]} observed the saturation
effect in the $sp^{2}-sp^{3}$ configuration ratio with the exposure dosage
growth, with the essential prevailing of $sp^{2}$-type configurations. It
seems reasonable to suppose that the cluster size is to decrease down to a
single aromatic ring at intensive bombarding. It was actually shown in
Ref. [\ref{7}] that disorder greatly reduce the probability of clustering,
with single $6$-fold distorted rings becoming the most probable configuration.
The local density-of-states function was calculated for different types of
ring distortion, taking $\sigma -\pi $ mixing due to out-of-plane distortion
of the ring into account. It was shown that the typical value of $E_{g}\sim
1 $ eV, which is characteristic for amorphous carbon, can be obtained for a
single $6$-fold ring with a ``chair'' distortion. The role of a ring
distortion was also discussed by Lee {\it et al.} [\ref{11}].

In this paper we propose a microscopic model in the framework of which we
managed to associate the properties of $sp^{2}$ electronic configuration
inherent to the stable carbon state with the local potential relief
peculiarities of a-C.

Let us first summarize the bonding possibilities of a carbon atom placing
the primary emphasis on the interrelation between the local crystalline
structure and the electronic band peculiarities. As graphite is the stable
allotrope of carbon, many disordered forms of carbon have structures based
on its lattice. The key feature of the electronic structure of amorphous
carbon is that it possesses a narrow optical gap of $0.4-0.7$ eV [\ref
{Robertson(Adv.Phys.1986)}], which is in contrast to the slight band overlap
in graphite and the $5.5$ eV band gap of diamond. Now let us suppose that
some carbon atoms of graphite are allowed to displace perpendicular to the
$\sigma $ bonding plane of graphite structure. As a matter of fact, by doing
so we change the local symmetry of bonding and introduce the $sp^{3}$-type
features into the $sp^{2}$ hybridization. We can anticipate the band gap
opening with such a lattice distortion as being peculiar to the $sp^{3}$
hybridization [\ref{Compagnini}]. In the absence of a long-range order the
energy fluctuation leading to the above mentioned distortion seems to be
more probable. The microscopic mechanisms that can be responsible for the
stabilization of such a distortion are well-known in the theory of
ferroelectric and superionic states. In accordance with the vibronic
mechanism [\ref{Bersuker}-\ref{Kristoffel}], the ferroelectric distortion
originates from the lattice instability due to the strong interaction
between the dipole moment of virtual interband electronic transition and
some transverse optical phonon. Such an interaction results in a phonon
softening, with the phonon frequency approaching zero at the transition
point.

The other possible mechanism suggests a strong hybridization of electronic
states of different symmetry at the top of the valence band [\ref{Cohen}-\ref
{our1}]. The escape of an electron from one of these states leaves a hole.
Incomplete screening (for symmetry reasons) of the hole results in the
appearance of an effective electron-hole dipole. The interaction of this
dipole with a transverse optical phonon can give rise to the formation of a
local potential minimum additional to the lattice site, {\it i.e.}, to the
double-well (DW). The essential energy overlap of $\sigma $-type and
$\pi $-type states within the valence band, which takes place in the amorphous
carbon, allows us to believe the latter mechanism to be more suitable for
this system. The spacing of potential minima of a double-well was found to
be [\ref{our}] 
\begin{equation}  \label{eqn:100}
\delta \simeq f_{exc}\left[ \frac{2}{M_{i}\bar{\omega}^{2}}A\right] ^{1/2},
\end{equation}
where 
\[
A=\frac{1}{N}\sum_{q}|\gamma _{q}|^{2}\omega _{q} 
\]
is the eigenenergy shift due to the electron-hole-phonon coupling. $M_{i}$
is the effective ionic mass, $N$ is the total number of ions, $\bar{\omega}$
is the characteristic phonon frequency ($\hbar =1$), $\gamma _{q}$ is the
electron-hole pair$-$TO phonon coupling constant, and $f_{exc}$ is the
expectation value for the operator of number of electron-hole excitations.
The evaluation of the lattice deformation energy yields [\ref{our1}] 
\begin{equation}  \label{eqn:2}
A\sim \frac{V^{2}}{M_{i}\overline{\omega }^{2}a^{2}},
\end{equation}
at low ($T\ll T_{D}$) and 
\[
A\sim 2\frac{V^{2}}{\overline{\omega }}\frac{T}{M_{i}\overline{\omega }%
^{2}a^{2}}, 
\]
at high ($T\gtrsim T_{D}$) temperatures ($k_{B}=1$). Here $T_{D}$ is the
Debye temperature, $a$ is the bond length, and $V$ is the characteristic
potential felt by an electron-hole pair ($V\sim E_{g}$).

The value of $f_{exc}$ is estimated as [\ref{our1}] 
\[
f_{exc}\sim \exp \left\{ -\frac{\varepsilon _{c}-V}{T}\right\} , 
\]
where $\varepsilon _{c}$ is the characteristic energy of electrons having
escaped the valence $\pi $ and $\sigma $ states with the hole formation. The
evaluation made with the following set of parameters $M_{i}\approx 2\times
10^{-23}$ g, $\bar{\omega}\approx 1300$ cm$^{-1}$, $a=1.4$ \AA , and $%
E_{g}\simeq 0.4-0.7$ eV yields, as the first approximation, $f_{exc}\sim 1$
and the ion displacement $\delta \simeq 0.2-0.35$ \AA . This finding is in
line with the result of calculations presented in Ref. [\ref{7}], where the
optical gap of a-C was calculated, with the ion out-of-plane displacement
value taken as the starting point.

The modification of potential shape with, {\it e.g.}, temperature increase
is mainly due to the DW spacing $\delta $ growth (Eqs. (\ref{eqn:100})-(\ref
{eqn:2})) as well as due to the difference in minima depth change which
reads as follows [\ref{our}] 
\begin{equation}  \label{eqn:3}
\Delta _{\varepsilon }\simeq \varepsilon _{c}-\Delta W-A/2,
\end{equation}
where 
\begin{equation}  \label{eqn:0}
\Delta W=W\left( 
\begin{array}{cccc}
l & l & l & l \\ 
1 & 2 & 2 & 1
\end{array}
\right) -W\left( 
\begin{array}{cccc}
l & l & l & l \\ 
1 & 2 & 1 & 2
\end{array}
\right)
\end{equation}
is the difference between the one-site $\pi -\sigma $ electron direct and
exchange Coulomb interaction terms [\ref{Haken}]. It should be noticed here
that the displacement of carbon ion from the basal $\sigma $-plane is
associated with the distortion of chemical bonds such that the $z$-component
of $\sigma $-like bonds appears to overlap with the $\pi $ electron state.

The potential evolution is manifested in changes of lattice dynamics. The
dispersion equation of the dielectric function includes the DW-associated
contribution of the form [\ref{Rome}] 
\begin{equation}  \label{eqn:12}
\epsilon ^{*} \propto \frac{1}{\int \exp \left\{ -\frac{V(x)}{T}\right\} dx}%
\int \frac{\exp \left\{ -\frac{V(x)}{T}\right\} dx}{M_{i}\omega
^{2}-iM_{i}\gamma \omega -\nabla ^{2}V(x)},
\end{equation}
where $V(x)$ is of the double-well type.

In order to simulate the potential relief we have to specify some trial
function $V_{tr}$ possessing desired properties. The function 
\[
V_{tr}=-T\ln \left[ \exp \left\{ -\frac{V_{1}}{T}\right\} +\exp \left\{ -%
\frac{V_{2}}{T}\right\} \right] , 
\]
where 
\[
V_{1}=-E_{1}+\frac{1}{2}kx^{2}, 
\]
\[
V_{2}=-E_{2}+\frac{1}{2}k(x-\delta )^{2}, 
\]
seems to be adequate enough to reproduce the principal features of the true
local potential, provided that $E_{1}-E_{2}=\Delta _{\varepsilon }$ and
$k=M_{i}\bar{\omega}^{2}$. With allowance made for conservation of the number
of particles within the elementary cell, which is possible under the
condition of neglecting any long-distance diffusion processes, $\delta $,
$\Delta _{\varepsilon }$, and $\overline{\omega }$ parameters completely
determine the relief properties. The analysis reveals a new feature of
spectrum which appears at [\ref{Rome}] 
\begin{equation}  \label{eqn:4}
\omega _{eff}\simeq \bar{\omega}\left| \sqrt{1-\frac{M_{i}\bar{\omega}%
^{2}\delta ^{2}}{4T}}\right|
\end{equation}
as additional to the bar oscillation $\bar{\omega}$. In accordance with
Eq.~(\ref{eqn:4}), the interwell spacing $\delta $ increase results in
the mode softening with its possible change from the oscillator- to the
relaxator type when $\delta >\frac{2}{\bar{\omega}}\sqrt{T/M_{i}}$.
The maximum of the
appropriate contribution to the imaginary part of the dielectric function,
{\it \ i.e.}, the maximum of absorption, is found to be at $\omega _{eff}$
for the oscillator and at $\omega _{eff}/\sqrt{3}$ for the relaxation-type
motion. Substitution of $\delta =0.2$ \AA\ into Eq. (\ref{eqn:4}) yields
$\omega _{eff}\simeq 670$ cm$^{-1}$ (oscillation-like), whereas
$\delta =0.35$ \AA\ corresponds to a relaxation-like motion,
with the absorption maximum
being near $840$ cm$^{-1}$ at room temperature.

These values of vibration mode frequencies are only estimates.
Nevertheless, in the framework of the model discussed it is possible
to interpret the appearance of weak vibration bands around
800 cm$^{-1}$ in irradiated graphite and diamond-like carbon
[\ref{lee93}], sputtered amorphous carbon [\ref{wang93}] as well as
glassy carbon [\ref{fli92}]. It should also be noted that
small but well defined {\it low frequency} shift ($\approx$ 80 cm$^{-1}$) of
G-band has experimentally been observed upon going from graphite
to diamond-like carbon films [\ref{yosh88},\ref{dil84}].
                    
In conclusion, we have suggested the model which allows one to associate the
structural peculiarities of amorphous carbon with the electronic spectrum
features. It should be mentioned that applicability of the above model is
restricted to the initial stage of the instability of $sp^{2}$-coordinated
state. The model reveals the microscopic source of instability in specific
electron-phonon interaction which results in the DW local potential
appearance. The modification of a single DW ignoring the inter-DW
interaction falls within the model limits as well.

At first glance it would seem that our approach is in line with the
Robertson's ideas [\ref{7}]. However, the DW potential we introduced is
inherently dynamical, {\it i.e.}, a carbon atom displaced from the
equilibrium $sp^{2}$ site towards the position out of plane will inevitably
return to the initial point. The characteristic residence time corresponding
to a carbon localized in the equilibrium $sp^{2}$ site ($\tau _{eq}$) far
exceeds that peculiar to the displaced position ($\tau _{d}$), $\tau _{eq}$ $%
\gg \tau _{d}$. From this point of view the percentages of fourfold- and
threefold-coordinated atoms are merely the $\tau _{d}/(\tau _{eq}+\tau _{d})$
and $\tau _{eq}/(\tau _{eq}+\tau _{d})$ ratios. Thus, as far as the
characteristic time ratio depends on the temperature as well as on the
sample preparation and post-growth treatment conditions, the model enables
one to consistently interprete available experimental results.

{\bf {Acknowledgments.}} A.P.L. gratefully acknowledges the support
by the ARPA MDA-90-J-1001,
The State of Texas through the Texas Center for Superconductivity
at the University of Houston (TCSUH) and the NSF through the
Material Science and Engineering Center at the University of
Houston (MRSEC-UH).

\end{document}